\documentclass[12pt]{article}
\usepackage[latin1]{inputenc}
\usepackage{amsfonts}
\usepackage{amssymb,amsmath}

\def\D{{\cal D}}

\def\NP{{\it Nucl. Phys.}}

\def\PL{{\it Phys. Lett.}}
\def\PR{{\it Phys.  Rev.}}

\def\MPL{{\it Mod. Phys. Lett.}}
\def\JHEP{{\it J.High\ Energy\ Phys.}}
\def\RMF{{\it Rev.\ Mex.\ Fis.}}

\title{Interacting D2-branes in 10 dimensions and non
abelian Born-Infeld theory} \vskip 5.5cm
\author{R.Gianvittorio, A. Restuccia, J. Stephany}
\date{}

\begin{document}
\maketitle \thispagestyle{empty} \vspace{-7cm} \hfill{Preprint
{\bf SB/F/05-334}} \hrule \vspace{5.5cm}
\begin{center}
\textit{Universidad Sim{\'o}n Bol{\'\i}var, Departamento de F{\'\i}sica\\
Apartado 89000, Caracas 1080A, Venezuela.}\vspace{1.cm}\\
\textit{ritagian@usb.ve, arestu@usb.ve, stephany@usb.ve}
\end{center}
\begin{abstract}
In this paper we extend the bosonic $D$-brane action in $D=10$
obtained by duality from the $D=11$ membrane wrapped on $S^1$ to
an $SU(2)$ non abelian system. This system  presents only first
class constraints,  whose  algebra closes off-shell and
generalizes the algebra of diffeomorphisms of the $D2$-brane to
include non abelian symmetry generators.

From the $SU(2)$ $D$-brane action, we also obtain the SU(2)
Born-Infeld theory by performing a covariant reduction to a flat
background. This calculation  agrees up to fourth order with the
result obtained from the superstring amplitudes and gives an
alternative approach to analyze non-abelian Born-Infeld theories.
\end{abstract}

\vskip 2.5cm \hrule
\bigskip
\centerline{\bf UNIVERSIDAD SIM{\'O}N BOL{\'I}VAR} \vfill

\section{Introduction}

The abelian Born-Infeld ($BI$)\cite{Fradkin85,Leigh89,Tseytlin96}
action was derived as the  part of the open string effective action
in an abelian  background  which depends on the field strength but
not on its  derivatives \cite{Fradkin85}. The $BI$ action appears
also as the effective action for the vector field on the world
volume of a $D$-brane \cite{Leigh89,Tseytlin96} where the induced
metric is taken to be Minkowskian.

For $n$ parallel branes the gauge invariance group is extended to
$U(1)^n$ which in the limit of overlapping branes is promoted to
$U(n)$ \cite{Witten96}. In analogy with the abelian case the
non-abelian Born-Infeld ($NBI$) action should be the string
effective action in a non-abelian background or  the effective
action of the vector field in the $n$-brane configuration. But in
this case the part of the  action which depends on the field
strength of the non-abelian vector field but not on its covariant
derivatives is not unambiguously defined. This is because the
relation $\left[ \D_m,\D_n \right] F_{pq} = \left[ F_{mn} , F_{pq}
\right]$ allows to transform some terms with covariant derivatives
into terms without derivatives. In Refs.
\cite{Tseytlin97,Tseytlin99} Tseytlin suggested a natural
prescription for defining the $NBI$ action obtained by applying a
symmetrized trace to the adjoint gauge group indices in the action
defined by the replacement of $ F_{mn}$ in the $BI$ action by a
non abelian field strength (see Ref. \cite{Hagiwara81} for an
early discussion). With this prescription, the $NBI$ lagrangian
does not contain commutators of $F$'s as a result of being
completely symmetric in all factors of $F$ in each monomial $tr
\left( F...F\right)$. The proposal reproduces exactly the full
non-abelian open string effective action up to order four
\cite{Dai89} since all other possible terms with covariant
derivatives can be redefined away at this order. At order $F^6$ it
was pointed in Ref.\cite{Hashimoto97} that Tseytlin's $NBI$ fails
to reproduces the fluctuations spectrum computed in perturbative
string theory \cite{Berkooz96}.

More recently a different proposal to fix the ambiguities of the
$NBI$ action requiring the existence of certain BPS configurations
was proposed in Ref. \cite{De Fosse01} and used to compute the
action up to fifth order \cite{Koerber01,Koerber01b} and sixth
order \cite{Koerber02}. The $F^5$ result was obtained in
Ref.\cite{Medina02}  from the open superstring 5-point amplitude
and confirmed by an independent method in \cite{Grasso02}. In
Refs.\cite{Sevrin03,Nagaoka03} the $F^6$ result was compared
favorably with the string theory computations but no definite
expression for the effective action up to sixth order has been
established yet.

In order to analyze the NBI theory it is natural to work in 10
dimensions and to look at its origin on the superstring
amplitudes. From the NBI in 10 dimensions one may then obtain all
non-abelian $D$-brane actions, in a particular gauge, in lower
dimensions by dimensional reduction.

Another approach to this problem is to start from the membrane or
supermembrane in 11 dimensions and  obtain the geometrical NBI
structure by considering the $D=11$ object compactified on $S^1$.
The dual to this systems is a $D2$ brane in 10 dimensions which is
manifestly invariant under the diffeomorphisms on the world volume.
In each case   the gauge field are of course abelian. The actions
for the $D2$ brane obtained in this way  agree both in the bosonic
and in the supersymmetric case with the Dirac Born Infeld actions
when a perturbative expansion on the field strength is valid.
\cite{PKTow,CScmid,Ovalle00,Ovalle00b}.

It is accepted that two interacting $D$-branes in the limit when
they coincide should be described in terms of $SU(2)$ gauge fields
together with the transverse coordinates which also carry an index
in the adjoint representation of the group and could be
interpreted as noncommutative coordinates \cite{Witten96}. It can
be argued that for consistency the system must be invariant under
diffeomorphisms besides having the $SU(2)$ gauge symmetry. In
order to describe this situation one may consider all possible
extensions of the diffeomorphisms algebra which include the
generators of $SU(2)$ gauge symmetry. If the extended algebra is
realized in terms of first class constraints then there is a
unique way to construct the hamiltonian of the theory, i.e  just
to take the linear combination of the first class constraints
(this is so for any system invariant under diffeomorphisms).
Finally if that hamiltonian arises as the canonical formulation of
a covariant lagrangian one has succeeded in constructing a
representation of the interacting system. We present in this work
a description of this procedure in the low energy regime where the
transverse coordinates behave as commutative ones, while the gauge
fields are non-abelian.

The main point in the construction is the extension of the
diffeomorphisms algebra to include the generators of the $SU(2)$
gauge symmetry. The realization is done using a set of first class
constraints expressed in terms of the commutative transverse
coordinates and the $SU(2)$ gauge fields. This realization in
terms of these fields is unique. There could be other realizations
of the algebra but they necessarily include other fields than the
commutative coordinates and gauge fields. There exists also a
covariant Lagrangian whose canonical hamiltonian is the unique one
associated to the extended algebra.

In order to construct from this theory a NBI action we fix the gauge
and reduce the theory to a flat background. The NBI action in 2+1
dimensions emerges as a perturbative expansion in terms of the
curvature of the non abelian gauge field. We show the explicit
calculations, for $SU(2)$, till the sixth order on the curvature.
The terms up to fourth order without derivatives or commutators of
the curvatures may be rewritten in terms of a symmetrized trace of
the Born-Infeld action. This is so because classical $D=11$
supermembrane action reproduces the  sigma-model one-loop
contribution of the superstring Ref.\cite{PKTow}.

The approach presented here appears to be generalizable to the
supersymmetric case \cite{Bergshoeff01,DeRoo98,De Roo02} where
issues related to the $\kappa$-symmetry and the non commutativity
of the embedding maps introduce different obstacles for the
construction of the action.  See also Ref.\cite{De Boer03,De
Boer03b} for a different approach for defining the $NBI$ action
based in general coordinate invariance.

\section{Non-abelian interacting $D2$ branes}

In $d$ dimensions the $D2$-brane is described  in terms of the
intrinsic three dimensional metric $\gamma^{ij}$, the space time
coordinates $X^m$ and a gauge field $A_i^I$ defined on the world
volume\cite{Ovalle00,Ovalle00b,Khoudeir98}. A natural  extension
for the action in the non abelian case is given by,

\begin{eqnarray}
S(\gamma,X,A)&=& -\frac{1}{2}\int
d^3\xi\sqrt{-\gamma}\left(\gamma^{ij}\partial_iX^m\partial_jX^n\eta_{nm}
+\frac{1}{2}\gamma^{ij}\gamma^{kl}F_{ik}^IF_{jl}^I-1\right);\hspace{0.5cm}\\
m &=& 0,...,d-1 \nonumber  \label{cov}
\end{eqnarray}
where $F$ is the curvature of the $SU(2)$ connection $A$:
\begin{equation}
F_{ij}^I=\partial_iA_j^I-\partial_jA_i^I + f^{IJK}A_i^JA_j^K.
\end{equation}
The dimension $d$ of the target space is not relevant in the
bosonic case although it would be in the supersymmetric extension.
The canonical analysis may be done using the ADM decomposition of
$\gamma^{ij}$ given by,
\begin{equation}
\begin{tabular}{l}
$\gamma _{ab}=\beta _{ab} \hspace{3.2cm}\gamma ^{ab}=\beta^{ab}-N^{a}N^{b}N^{-2}$ \\
$\gamma_{00}=-N^2+\beta_{ab}N^aN^b\hspace{1.2cm}\gamma ^{00}=-N^{-2}$\\
$\gamma ^{0a}=N^{a}N^{-2}\hspace{1.2cm}\gamma_{0a}=\beta_{ab}N^b
\hspace{1.2cm} a,b=1,2$ .
\end{tabular}
\label{adm}
\end{equation}
where
\begin{equation}
\beta ^{ab}\beta _{bc}=\delta _{c}^{a};\hspace{0.8cm}
\sqrt{-\gamma }=N\sqrt{\beta }, \ \ \beta=det\beta_{ab} .
\end{equation}
In terms of these variables the action is given by,
\begin{eqnarray}
S(\beta,X,A)=-\frac{1}{2}\int d^3\xi N
\sqrt{\beta}\left(-N^{-2}\dot{X}^m\dot{X}_m+2N^aN^{-2}\dot{X}^m\partial_aX_m\right.
\nonumber \\  +
(\beta^{ab}-N^aN^bN^{-2})\partial_aX^m\partial_bX_m
-1+\frac{1}{2}(\beta^{ac}\beta^{bd}-2\beta^{ac}N^bN^dN^{-2})F_{ab}^IF_{cd}^I\nonumber\\
 \left.-N^{-2}\beta^{ab}F_{0a}^IF_{0b}^I +
2N^{-2}\beta^{ac}N^bF_{0a}^IF_{bc}^I\right);\hspace{2.0cm}m=0,...,d-1\
. \label{dbbf}
\end{eqnarray}
Introducing the canonical momenta $P_m$, $\Pi^{Ia}$  and $P_{ab}$
associated to $X^m$, $A_a^I$ and $\beta^{ab}$ with the canonical
algebra
\begin{eqnarray}
\{X^{m}(\xi),P_{n}(\xi')\}=\delta^{m}_{n}\delta^2(\xi-\xi'),\nonumber\\
\{A_a^I(\xi),\Pi^{Jb}(\xi')\}=\delta^{IJ}\delta^{b}_a\delta^2(\xi-\xi') \\
\{\beta^{ab}(\xi),P_{cd}(\xi')\}=\frac{1}{2}
(\delta^a_c\delta^b_d+\delta^b_c\delta^a_d)\delta^2(\xi-\xi')\nonumber
\end{eqnarray}
and after some algebra we obtain the hamiltonian of our
interacting $D2$-brane system  expressed in the form,
\begin{equation}
H=\int
d^2\xi\left(\Lambda\Phi+\Lambda^a\Phi_a+A_0^I\varphi^I+\Sigma^{ab}\Omega_{ab}\right)
\label{hcero}
\end{equation}
where $\Phi$, $\Phi^a$, $\varphi^I$ and $\Omega_{ab}$ are the
constraints associated to the symmetries of the $SU(2)$ interacting
$D2$-brane. The constraints are given by,
\begin{equation}
\Omega_{ab}\equiv P_{ab}\approx0\label{vin1}
\end{equation}
\begin{eqnarray}
\Phi &\equiv & \left(P^{m}P_{m} + \Pi^2 + \beta \beta^{ab}
\partial _{a}X^{m}\partial _{b}X^{m} - \beta + f^I f^I \right)\approx0  \nonumber\\
 \label{vinf}
\end{eqnarray}
\begin{equation}
\Phi_a\equiv\left(
\partial _{a}X^{m}P_{m}+\Pi^{Ib}F_{ab}^I\right)\approx0\label{vinfa}
\end{equation}
\begin{equation}
\varphi^I \equiv  - \left(\D
_{a}^{IJ}\Pi^{Ja}\right)\approx0\label{varfi}
\end{equation}
where,
$$
F_{ab}^I \equiv \varepsilon_{ab} f^I, \ \ \ \ \D_a^{IJ}  \equiv
\partial_a \delta^{IJ} + f^{IKJ} A_a ^K
$$
and
$$\Pi^2  \equiv  \beta_{ab} \Pi^{Ia} \Pi^{Ib}$$
These constraints generalize the primary constraints of the
abelian case. The conservation of the primary constraints give
rise to the secondary constraint,
 \begin{equation}
\label{condition} \frac{\delta \Phi} {\delta\beta^{ab}}=0 ,
\end{equation}
which allows to determine $\beta^{ab}$ in terms of the other
canonical variables. The elimination of the canonical pair
$\beta^{ab},P_{ab}$ allows the reduction of the action. The only
constraint which depends on $\beta^{ab}$ is $\Phi$.  All the
constraints are now first class ones: $\Phi_a $ are the generators
of spatial diffeomorphisms on the world volume and they do not
depend on the target metric nor on the world volume metric
$\beta^{ab}$ and in this sense they are topological; $\Phi$ is the
generator of temporal diffeomorphisms and is metric dependent.
Because of (\ref{condition}), the algebra satisfied by (\ref{vinf}),
(\ref{vinfa}) and (\ref{varfi}) is the same whether we replace the
explicit expression of $\beta^{ab}$ or consider it as an independent
variable. The algebra satisfied by $\Phi_a$, $\Phi$ and $\varphi^I$
is
\begin{eqnarray}
\{ \Phi(\xi), \Phi(\xi') \}&=& \left( C^{ab} \Phi^a(\xi) +  C'^{ab} \Phi _a(\xi') \right) \partial_b \delta^2(\xi-\xi') .\nonumber \\
\{\varphi ^{I}(\xi) , \varphi ^{'J}(\xi')\}&=&f^{IJK} \varphi_K \delta^2(\xi-\xi') \nonumber \\
\{ \Phi_a(\xi'), \Phi_b(\xi')\}&=& \Phi_a(\xi') \partial_b
\delta^2(\xi-\xi')+ \Phi_b(\xi) \partial_a
\delta^2(\xi-\xi') - \varepsilon_{ab} f^J \varphi^J(\xi) \delta^2(\xi-\xi').\nonumber \\
\{\Phi(\xi) , \varphi ^{I}(\xi')\}&=&0 , \  \  \  \  \   \  \  \{\Phi_a(\xi) , \varphi ^{I}(\xi')\}=0 .\nonumber \\
 \{\Phi_a(\xi) , \Phi(\xi')\} &=&\left( \Phi(\xi) + \Phi(\xi') \right)  \partial_a  \delta  + C_a ^I \varphi^I \delta^2(\xi-\xi').
\end{eqnarray}
where $C^{ab}= 4 \beta \beta^{ab}$ and $C_a ^I$ is a function of $
(\beta^{ab}) $. Since the structure coefficients depend on
$\beta^{ab}$ the algebra is open. We have thus constructed the
canonical formulation associated to the original covariant action
(\ref{cov}).

The expression for the Hamiltonian is closed if a solution of
Eq.(\ref{condition}) is provided. In the next section we obtain a
perturbative solution of Eq.(\ref{condition}).

We may consider now the inverse problem. Let us start from the
constraints (\ref{vinfa}) and (\ref{varfi}) which generate the
spatial diffeomorphisms and $SU(2)$ gauge symmetry respectively
together with (\ref{vin1}) which express the fact that in the
action there are no time derivatives of the auxiliary metric
$\beta_{ab}$. The expression  (\ref{vinfa}) is unique since it
must be topological. The expression (\ref{varfi}) is the usual
Gauss constraint for a gauge field. If we ask for an extension of
the abelian expression for $\Phi$ which together with
(\ref{vinfa}) and (\ref{varfi}) closes an algebra, we obtain a
unique solution given by (\ref{vinf}). The hamiltonian is then
also uniquely defined and given by the linear combination of these
first class constraints. Finally one may recognize the covariant
action (\ref{dbbf}) as the one whose canonical formulation give
rise to the hamiltonian \ref{hcero}. This give thus a strong
justification to consider the action (\ref{dbbf}) as the starting
point.

\section{The computation of $\beta_{ab}$}

In order to construct the action explicitly we have to determine the
solution for $\beta$. This can be done  order by order in an
expansion in terms of the momenta of the gauge fields. First we
compute (\ref{condition})  obtaining,
\begin{equation}
\Pi^{Ia}\Pi^{Ib}+\beta \beta^{ab} \beta^{cd} \partial _{c}X^{m}
\partial _{d}X^{m} - \beta  \beta^{ab} - \beta  \beta^{ac} \beta^{bd}
\partial _{c}X^{m} \partial _{d}X^{m} =0,  \label{deltaphi}
\end{equation}
from where we get after contracting with $\beta_{ab}$
\begin{equation}
\Pi^{Ia}\Pi^{Ib}+\beta \beta^{ab}  - \beta^{ab}  \Pi^2 - \beta
\beta^{ac} \beta^{bd} \partial _{c}X^{m} \partial _{d}X^{m} =0.
\label{deltaphi2}
\end{equation}
After some calculations the determinant of (\ref{deltaphi2}) may be  obtained. For $\beta \neq 0$ we have
\begin{equation}
det \left( \Pi^{I\bullet}\Pi^{I\bullet} \right) +\beta - \Pi^2 =
det\left( \partial _{\bullet}X^{m} \partial _{\bullet}X^{m} \right)
, \label{deltaphi3}
\end{equation}
where the $\bullet$ means a target space index $a=1,2$. From
(\ref{deltaphi2}) and (\ref{deltaphi3}) , and provided $\beta \neq
0$ we have
\begin{eqnarray}
& &\Pi^{Jc}\Pi^{Jd}\Pi^{Ia}\Pi^{Ib} \beta_{ac}\beta_{bd}+  \Pi^2
\left[ det\left( \partial _{\bullet}X^{m}
\partial _{\bullet}X^{m}  \right)  - det \left( \Pi^{I\bullet}\Pi^{I\bullet} \right) \right]  \nonumber \\
& & - \beta  \left( \Pi^{Jc}  \partial _{c}X^{m} \right)   \left(
\Pi^{Jd}  \partial _{d}X^{m} \right) =0. \label{deltaphi4}
\end{eqnarray}
We finally obtain,
\begin{equation}
\beta^{ab} \left[ det\left( \partial _{\bullet}X^{m} \partial
_{\bullet}X^{m} \right)  - det \left( \Pi^{I\bullet}\Pi^{I\bullet}
\right) \right] + \Pi^{Ia}\Pi^{Ib} - \beta\beta^{ac} \beta^{bd}
\partial _{c}X^{m}
\partial _{d}X^{m}  =0 \label{deltaphi5}
\end{equation}
which allows to solve  $\beta_{ab}$ order by order  in
$\Pi^{Ia}\Pi^{Ib}$  using an iterative procedure. We also realize
that the constraint $\Phi$ only depends on  $\beta_{ab}$ through
$\beta$ since we can express
\begin{equation}
\Phi = P^{m}P_{m} + f^I f^I +  \beta
\end{equation}
Thus we need only to evaluate $\beta$. In the abelian case $det
\left( \Pi^{\bullet}\Pi^{\bullet}\right)=0$ and the  exact solution
for $\beta$ is given then by,
\begin{equation}
\beta=det\left( \partial _{\bullet}X^{m} \partial _{\bullet}X^{m}
\right) + \Pi^{c}  \Pi^{d} \partial _{c}X^{m} \partial _{d}X^{m} .
\end{equation}
The exact closed expression for the constraint $\Phi$ for the
abelian case is
\begin{equation}
\Phi \equiv  P^{m}P_{m} + f f + det\left( \partial _{\bullet}X^{m}
\partial _{\bullet}X^{m} \right)  +\Pi^{c}  \Pi^{d} \partial
_{c}X^{m} \partial _{d}X^{m} .
\end{equation}

Returning to the non-abelian case  we try a power  expansion for
$\beta_{ab}$
\begin{equation}
\beta_{ab} =g_{ab} + O_{1 ab} + O_{2 ab} + ....  \ \ ,
\end{equation}
with  the induced metric $g_{ab} \equiv \partial _{a}X^{m}
\partial _{b}X^{m} $,  and taking $O_{n ab}$ of order $2n$ in
$\Pi^{Ia}$. We then find for $\beta^{ab}$
\begin{equation}
\beta^{ab} =g^{ab} - g^{ac}  O_{1 cd} g^{db} - g^{ac}  O_{2 cd}
g^{db} + g^{ac}  O_{1 ce} g^{ef} O_{1 fd} g^{db}+ ....  \ \ ,
\end{equation}
and for $\beta$,
\begin{equation}
\beta =g + \left(\Pi_g\right)^2 -   det \left( \Pi^{I
\bullet}\Pi^{I \bullet} \right) + O_{1 ab}  \Pi^{Ia} \Pi^{Ib} +
O_{2 ab} \Pi^{Ia} \Pi^{Ib} + ... \label{beta}
\end{equation}
where $g=det\left(g_{ab}\right)$ and $\left(\Pi_g\right)^2=g_{ab}
\Pi^{Ia} \Pi^{Ib}$. After some calculations we obtain the solution
for the $O_{i ab}$ to order $i=2$ ,

\begin{eqnarray}
g O_{1 ab}  & = & g_{ab}  \left(\Pi_g\right)^2 - g_{ac}  g_{bd} \Pi^{Ic}  \Pi^{Id}  \label{O1} \\
g O_{2 ab}  & = & g_{ab}   O_{1 cd}  \Pi^{Ic}  \Pi^{Id}  + 2 g
O_{1 ac} g^{cd} O_{1 bd} - 2 g_{cd}  \Pi^{Ic}  \Pi^{Id} O_{1 ab}
\label{O2}
\end{eqnarray}
 In the abelian case $O_{1 ab}=0$ and  $O_{2 ab}=0$ and  $det
\left( \Pi^{\bullet}\Pi^{\bullet}\right)=0$ in agreement with the
previous result.

\section{Covariant Reduction of the interacting theory to a Non-abelian Born-Infeld Theory}
In this section we  perform a covariant reduction of the
interacting theory action previously obtained to a NBI action. We
start from the action
\begin{equation}
 S= \left<  P_m \dot{X}^m + \Pi^{Ia} \dot{A}^I_a - H \right> \label{action}
\end{equation}
where the hamiltonian is given by (\ref{hcero}). The field
equations are
\begin{eqnarray}
\dot{X}^m &=& 2 P^m \Lambda +   \Lambda^a \partial_a X^m  \label{xpunto} \\
\dot{A}^I_a & = & \D_a A^I_0 + \Lambda \frac{\delta \beta}{\delta
\Pi^{Ia} }+ \Lambda^b F_{ba}^I   \label{apunto} \ \ .
\end{eqnarray}

Now we  look for class of solutions to the constraints that are
obtained  when we impose $g_{ab}=\delta_{ab}$.  We found them by
considering,
\begin{eqnarray}
X^0 &=& \tau   \ \ \ \   \  \  X^a= \sigma^a  \ \  \ a=1, 2 \nonumber\\
X^m &=& 0   \ \ \ \  P_m=0 \  \  \  m \geq 3\ \ .
\end{eqnarray}
Substituting this in Eq. (\ref{xpunto}) we obtain  for the momenta
\begin{eqnarray}
2 P^0 \Lambda &=& 1  \\
2P_a \Lambda+\Lambda_a &=&0 \ \ .
\end{eqnarray}
After imposing $\Phi_a =0$ and $\Phi=0$ we have
\begin{eqnarray}
P_a + \varepsilon_{ab} f^I \Pi^{Ib}  =0   \\
P_0 ^2=   P_a ^2 + f^I f^I + \beta \ \ .
\end{eqnarray}
We replace  all these relations in (\ref{action}) and obtain
\begin{equation}
 S= \left<  P_0  + \Pi^{Ia} \dot{A}^I_a -  \Pi^{Ia} \D_a A^I_0\right>
 \label{action2}\ \ .
\end{equation}
From  (\ref{apunto}) we get $F_{0a}^I$ and replacing  into
(\ref{action2}) we have
\begin{equation}
 S= \left<   \Pi^{Ia} F_{0a}^I - P^0\right> \label{action3} \ \ ,
\end{equation}
where
\begin{equation}
  P^0= \left(   \beta + f^I f^I + f^I f^J \Pi^{Ia}
  \Pi^{Ja}\right)^\frac{\,1}{2}\ \ .
\end{equation}
We may now expand the action in powers of the non abelian field
strength $F$'s.  In order to do that we need to express the
momenta $\Pi_a^{I}$ in terms of the $F_{0a}^I$.For example to
obtain the action to the fourth order in $F$'s we proceed in a
recursive way, order by order, and we first write $\Pi_a^{I}$ up
to order three in the field strength
\begin{equation}
  \Pi^I_a = P^0 F_{0a}^I - f^I f^J F_{0a}^J - 2 F_{0a}^I F_{0b}^J F_{0b}^J
  +  \varepsilon_{ac} \varepsilon_{bd}F_{0b}^I F_{0c}^J F_{0d}^J + 2 F_{0a}^J F_{0b}^J F_{0b}^I + O^5(F), \label{pi3}
\end{equation}
but retaining  $P^0$  only up to order two. The expansion of $P^0$
to order fourth gives
\begin{eqnarray}
  P^0 & = & 1 +  \frac{1}{2}  F_{0a}^I F_{0a}^I + \frac{1}{2}  f^I f^I
  +\frac{1}{4} F_{0a}^I F_{0a}^I f^J f^J \nonumber \\
  & & - \frac{9}{8} ( F_{0a}^I F_{0a}^I )^2
  - \frac{1}{2}  ( F_{0a}^I f^I )^2 + \frac{3}{4}  \varepsilon_{ac} \varepsilon_{bd} F_{0a}^I F_{0b}^I F_{0c}^J F_{0d}^J
   \nonumber \\
  & &
  + \frac{3}{2} F_{0a}^I F_{0a}^K F_{0b}^I F_{0b}^K - \frac{1}{8}( f^I f^I )^2 + O^6(F), \label{p03}
\end{eqnarray}

The Lagrangian  obtained up to fourth order by replacing in
(\ref{action3}) (\ref{pi3}) and (\ref{p03}) is given by
\begin{eqnarray}
 L &=&-1 - \frac{1}{2} f^I f^I +  \frac{1}{2} F_{0a}^I F_{0a}^I -  \frac{3}{8} \left( F_{0a}^I F_{0a}^I
 \right)^2 \nonumber \\
 & & +  \frac{1}{8} \left(  f^I f^I \right)^2 - \frac{1}{2} \left( f^I F_{0a}^I \right)^2
 +  \frac{1}{4}  F_{0a}^I F_{0a}^I f^K f^K \nonumber \\
   & & +  \frac{1}{2}  F_{0a}^I F_{0b}^I F_{0a}^K F_{0b}^K
 + \frac{1}{4}  \varepsilon_{ac} \varepsilon_{bd} F_{0a}^I F_{0b}^I F_{0c}^J F_{0d}^J  \nonumber  \\
\label{lagrangian4}
\end{eqnarray}
After eliminating all the terms with commutators  of the kind
$\left[ \D,\D\right] F = \left[ F , F \right]$ one obtains the
Lagrangian for the NBI system up to the fourth order.

\begin{eqnarray}
 L &=&-1 - \frac{1}{2} f^I f^I +  \frac{1}{2} F_{0a}^I F_{0a}^I +  \frac{1}{8} \left( F_{0a}^I F_{0a}^I \right)^2 \nonumber \\
  & & +  \frac{1}{8} \left(  f^I f^I \right)^2 - \frac{1}{4} \left( f^I F_{0a}^I \right)^2  \nonumber \\
 . \label{lagrangian}
\end{eqnarray}
The result obtained is manifestly covariant  and can be written as
\begin{equation}
 L = -1 - \frac{1}{4} F_{ij}^I F^{Iij} +  \frac{1}{32}  F_{ij}^I
 F^{Jij} F_{kl}^I F^{Jkl}
 . \label{covlagrangian}
\end{equation}

This Lagrangian agrees with Ref. \cite{Tseytlin99} once the
dimensional reduction to $2+1$ is performed.

Following the same procedure we obtain the Lagrangian up to the
sixth order

\begin{eqnarray}
 L &=&-1 - \frac{1}{2} f^I f^I +  \frac{1}{2} F_{0a}^I F_{0a}^I +  \frac{1}{8} \left( F_{0a}^I F_{0a}^I \right)^2 \nonumber \\
  & & +  \frac{1}{8} \left(  f^I f^I \right)^2 - \frac{1}{4} \left( f^I F_{0a}^I \right)^2  \nonumber \\
  & & +  \frac{1}{16} \left( F_{0a}^I F_{0a}^I\right)^3- \frac{3}{16} F_{0a}^IF_{0a}^I \left( f^J F_{0b}^J
  \right)^2 \nonumber \\
  & & + \frac{3}{16} f^J f^J \left( f^I F_{0a}^I \right)^2 - \frac{1}{16} \left(  f^I f^I
  \right)^3
  \label{lagrangian}
\end{eqnarray}

There is no general argument implying that this classical result
should agree to the sigma-model higher-loop contribution in string
theory, however we thing that the quantum deformation of the
algebra we have obtained may provide the complete result. \vskip
.5cm

\section{Discussion}

In this paper we showed that the bosonic $D$-brane action in $D=10$
obtained by duality from the $D=11$ supermembrane wrapped on $S^1$
may be extended to include $SU(2)$ non-abelian generators and that
the corresponding algebra of first class constraints is completely
consistent. This theory is formulated in terms of unconstrained
fields (i.e without second class constraints) and hence is at the
same level of consistency that the $U(1)$ $D$-brane theory. This is
achieved without including non-commutative coordinates. We thus
obtained a description of interacting $D$-branes theory in the low
energy regime where the transverse coordinates behave as commutative
ones, while the gauge fields are non-abelian.

From the $SU(2)$ $D$-brane action, we also obtained the SU(2)
Born-Infeld theory by performing a reduction to a flat background.
The calculations were performed till the sixth power of the
curvature and agree up to fourth order with the result obtained
from the superstring amplitudes. This is so because the
sigma-model one-loop calculation in the string theory is
reproduced by the classical supermembrane. There is no general
argument that this relation should be true for sigma-model
higher-loop calculation, however we thing that the quantum
deformation of the algebra of the first class constraints that we
have obtained should provide the complete result. It would be
interesting to have new computations of the effective action at
higher orders which could be compared with our results.

The complete understanding of interacting $D$-branes is still an
open problem. In particular, the non-abelian generalization of the
Born-Infeld theory still has unanswered  questions. One of them is
the role of the $\kappa$-symmetry in the supersymmetric
non-abelian $D$-brane theory. In this direction, the next problem
to be addressed  is the corresponding supersymmetric theory for
which our approach results very appropriate to analyze the role of
kappa symmetry.

\end{document}